\documentclass[aps,prd,twocolumn,nofootinbib]{revtex4-1}
\usepackage{epsfig}
\usepackage[colorlinks,linkcolor=blue,anchorcolor=blue,citecolor=blue,urlcolor=blue,breaklinks=true]{hyperref}
\usepackage{graphics}
\usepackage{slashed}
\usepackage{color}

\begin{document}
\author{Shu-Sheng Xu$^{1,3}$}
\author{Yan Yan$^{1}$}
\author{Zhu-Fang Cui$^{1,3}$}
\author{Hong-Shi Zong$^{1,2,3}$}\email{zonghs@nju.edu.cn}
\address{$^{1}$ Department of Physics, Nanjing University, Nanjing 210093, China}
\address{$^{2}$ Joint Center for Particle, Nuclear Physics and Cosmology, Nanjing 210093, China}
\address{$^{3}$ State Key Laboratory of Theoretical Physics, Institute of Theoretical Physics, CAS, Beijing 100190, China}

\title{2+1 flavors QCD equation of state at zero temperature within Dyson-Schwinger equations}
\begin{abstract}
Within the framework of Dyson-Schwinger equations (DSEs), we discuss the equation of state (EOS) and quark number densities of 2+1 flavors, that is to say, $u$, $d$, and $s$ quarks. The chemical equilibrium and electric charge neutrality conditions are used to constrain the chemical potential of different quarks. The EOS in the cases of 2 flavors and 2+1 flavors are discussed, and the quark number densities, the pressure, and energy density per baryon are also studied. The results show that there is a critical chemical potential for each flavor of quark, at which the quark number density turns to nonzero from 0; and furthermore, the system with 2+1 flavors of quarks is more stable than that with 2 flavors in the system. These discussion may provide some useful information to some research fields, such as the studies related to the QCD phase transitions or compact stars.

\bigskip


\bigskip

\noindent PACS Number(s): 21.65.Mn, 25.75.Nq, 21.65.Qr

\end{abstract}
\maketitle

\section{Introduction}\label{intro}
Quantum Chromodynamics (QCD) is commonly accepted as the basic theory of strong interaction. Because of the asymptotic freedom character of QCD, the high energy processes could be pictured hadrons as weakly interacting quarks and gluons. However, the interaction is so strong that the perturbation theory is invalid at energies below a few hundred MeV, and the picture is inadequate. Further more, confinement and dynamical chiral symmetry breaking (DCSB) are two fundamental properties of QCD in the region of low energy. On the one hand, the color confinement is not understand very well until now, it remains only an empirical fact that colored object have not been observed. On the other hand, DCSB is better understood, which explains the origin of constituent-quark masses and underlies the success of chiral effective field theory.

The study of QCD at nonzero temperature $T$ and finite baryon density $\rho_B$ is one of the most interesting topics in contemporary physics studies. It is believed that a phase transition happens in strongly interaction matter from a confined hadronic system at low temperatures and low densities to a deconfined quark gluon plasma (QGP) phase at high temperatures and/or densities.
There are several famous laboratories and experimental facilities which are related with thermal and dense QCD, such as the Relativistic Heavy Ion Collider (RHIC), the Large Hadron Collider (LHC), and the Facility for Antiproton and Ion Research (FAIR). RHIC and LHC are mainly concentrated on high temperature and low density physics, while FAIR focus on low temperature and high density plasma physics. It is known that the Equation of state (EOS) of QCD for cold and nonvanishing quark chemical potential $\mu$ plays a key role in the study of compact stars~\cite{ozel2006soft,alford2007astrophysics,klahn2007modern,PhysRevC.88.015802}. Since the perturbation QCD is invalid in the low-energy region, it is very difficult to obtain a reliable EOS from the first principles of QCD. A lot of models of neutron stars or quark stars have been studied~\cite{PhysRevD.9.3471,PhysRevC.61.045203,menezes2006quark}. Many important studies have been done in Refs. \cite{PhysRevD.82.101301,PhysRevD.80.103003}, and as data of masses and radii of neutron stars have accumulated, the EOS that best fits the observation data has been strictly constrained.
The calculations of the EOS~\cite{aoki2006equation,PhysRevD.75.094505,PhysRevD.77.014511,PhysRevD.85.094508,PhysRevD.90.094503,PLB.730.99} and the transition temperature~\cite{PhysRevD.71.034504,PhysRevD.74.054507,aoki2006qcd} based on the Lattice QCD can be performed with merely realistic quark mass spectrum. However, Lattice QCD confront sign problem, which refers to the difficulty of evaluating the integral of a highly oscillatory function numerically, when dealing with finite chemical potential. Consequently, nowadays various effective models are often and in some sense inevitably used to study related issues phenomenologically, such as the Nambu--Jona-Lasinio (NJL) model~\cite{PPNP.27.195,RMP.64.649,PR.247.221,PLB.591.277--284,PR.407.205,PRD.73.014019,TEPJC.49.213--217,PRD.77.114028,PLB.662.26,EPJC.73.2612,CUIJMP,EPJC.74.2782,PhysRevD.91.036006}, Lattice QCD~\cite{JHEP.04.050,JHEP.09.073,PhysRevLett.110.172001}, the Dyson-Schwinger equations (DSEs) ~\cite{PPNP.33.477,PPNP.45.S1,IJMPE.12.297,PPNP.61.50,JHEP.04.14,PPNP.77.1-69,JHEP.07.014,PhysRevD.90.114031,PhysRevD.91.034017,PhysRevD.91.056003}, and the Quantum Electrodynamics in (2+1) dimensions (QED$_3$)~\cite{PhysRevD.29.2423,PPNP.33.477,PhysRevD.90.036007,PhysRevD.90.065005,PhysRevD.90.073013}.
This study will mostly focus on the EOS at zero temperature and nonzero chemical potential, especially the region of high chemical potential, within the framework of DSEs. One of our main goal is to compare the stability of the dense system between 2 flavors (only the two lightest quarks are included, namely, $u$ and $d$ quarks) and 2+1 flavors (not only $u$ and $d$ quarks, but the slightly heavier $s$ quark is also included) cases.

In this paper, the quark number density in the case of 2+1 flavors is calculated by constrains of the chemical equilibrium and electric charge neutrality conditions within the framework of DSEs. The
Qin-Chang effective gluon propagator model~\cite{PhysRevC.84.042202}, which has been proved to be successive in hadron physics and thermal QCD recent years, is adopted to study the EOS in both 2 flavors and 2+1 flavors cases. The following of this paper is organized in such a way: In Sec.~\ref{dsesandgp}, we give a brief introduction to the DSEs at zero temperature as well as zero chemical potential first, and then fit the quark propagator using the formula with 3 complex conjugate poles. In Sec.~\ref{qndandpressure} we discuss in detail the quark number densities at $T=0$ algebraically, and show numerical results of the pressure in cases of 2 flavors and 2+1 flavors. Finally, we give a brief discussion and summary in Sec.~\ref{summary}.

\section{Dyson Schwinger equations at zero temperature and zero chemical potential}\label{dsesandgp}
In this section, we briefly introduce the Dyson Schwinger equations(DSEs) formalism, which are coupled integral equations relating the Green's functions for the theory to each other~\cite{PPNP.33.477,PPNP.45.S1,IJMPE.12.297,PPNP.61.50,PPNP.77.1-69}. The zero temperature and zero quark chemical potential version of the quark propagator DSE, which is the most important one among all the DSEs, reads
\begin{eqnarray}
S(p)^{-1} &=& Z_2(i\slashed{p}+Z_m m)\nonumber\\
&+& g^2 Z_{1F} \int_q \frac{\lambda^a}{2} \gamma_\mu S(q)\Gamma^a_\nu(p,q) D_{\mu\nu}(p-q),
\label{Eq:DSE1}
\end{eqnarray}
where $S(p)$ is the dressed quark propagator, $Z_2$ is the field-strength renormalization constant, $Z_m$ is the mass renormalization constant where $m$ is the current quark mass, $g$ is the coupling constant, $Z_{1F}$ is the quark-gluon-vertex renormalization constant, $\int_q :=\int \frac{d^4 q}{(2\pi)^4}$ is a symbol that represents a Poincar\'{e} invariant regularization of the four-dimensional Euclidean integral, $\lambda^a$ is the Gell-Mann matrices, $\Gamma^a_\nu(p,q)$ is the quark gluon vertex, and $D_{\mu\nu}(p-q)$ is the dressed gluon propagator. The general form of the quark propagator at zero temperature and zero chemical potential reads
\begin{equation}
S(p)^{-1}=i{\not\!p}A(p^2)+B(p^2),\label{quarkpg1}
\end{equation}
where $A(p^2)$ and $B(p^2)$ are scalar functions of $p^2$. The renormalization condition is
\begin{eqnarray}
A(\zeta^2)&=&1,\\
B(\zeta^2)&=&m,
\end{eqnarray}
at sufficiently large spacelike $\zeta^2$~\cite{PhysRevC.60.055214}. We choose the renormalization point to be $\zeta=19$ GeV in this work.
Eq.~(\ref{Eq:DSE1}) is the exact result from first principles of QCD, but we can not solve it directly unless concrete truncations are performed. Rainbow truncation is used in this work, which means a bare vertex is adopted
\begin{equation}
\Gamma^a_\nu(p,q)=\frac{\lambda^a}{2}\gamma_\nu\,,
\label{rainbow1}
\end{equation}
and the Qin-Chang gluon propagator model~\cite{PhysRevC.84.042202} is specified by a choice for the effective interaction in Landau gauge,
\begin{eqnarray}
g^2 D_{\mu\nu}(k)&=&\mathcal{G}(k^2) D^0_{\mu\nu}(k)\nonumber\\
&=&\frac{\mathcal{G}(k^2)}{k^2} P^T_{\mu\nu}(k),
\label{gluonmodel}
\end{eqnarray}
with
\begin{eqnarray}
P^T_{\mu\nu}(k)&=&\delta_{\mu\nu}-\frac{k_\mu k_\nu}{k^2},
\end{eqnarray}
the transverse projection operator, and
\begin{eqnarray}
\frac{\mathcal{G}(k^2)}{k^2}&=&\frac{8\pi^2}{\omega^4}D e^{-\frac{k^2}{\omega^2}} + \mathcal{F}_{UV}(k^2),
\\
\mathcal{F}_{UV}(k^2)&=&\frac{4\pi^2\gamma_m \mathcal{F}(k^2)}{1/2 \ln[\tau+(1+k^2/\Lambda_{QCD}^2)^2]},
\\
\mathcal{F}(k^2)&=&[1-e^{-k^2/(4m_t^2)}]/k^2,
\end{eqnarray}
where the parameters are $\gamma_m=12/(33-2N_f)$ with $N_f=4$, $\tau=e^2-1$ with $e=2.71828$, $\Lambda_{QCD}^{N_f=4}=0.234$ GeV, and $m_t=0.5$ GeV, which determined by behavior of ultraviolet of the gluon propagator. The remaining parameters $D$ and $\omega$ are fitted by the mass and decay constant of pion, we choose $\omega=0.7$, $D\omega=(0.8\mathrm{GeV})^3$ in this work as in Ref.~\cite{PhysRevC.84.042202}. Moreover, in our study the isospin symmetry is preserved in the strong interaction section, and the current quark mass $m_u=m_d=3.4$ MeV, $m_s=82$ MeV is used. The isospin symmetry will be broken when the electroweak interactions is included, which will be introduced in Sec.~\ref{qndandpressure} of this paper.

Now substitute Eqs.~(\ref{quarkpg1}), (\ref{rainbow1}), and (\ref{gluonmodel}) into Eq.~(\ref{Eq:DSE1}), and project Eq.(\ref{Eq:DSE1}) on the two tensor structures $T_1=i\slashed{p}$ and $T_2=\textbf{1}$, one can obtain the following equations for the two dressing functions $A(p^2)$ and $B(p^2)$,
\begin{eqnarray}
A(p^2)&=&Z_2+\frac{4}{3p^2}\int_q \frac{\mathcal{G}(k^2)}{k^2} \frac{A(q^2)}{q^2 A^2(q^2)+B^2(q^2)}
\nonumber\\
&&\hspace*{5mm}\times \left( p\cdot q +2 \frac{(k\cdot p) (k\cdot q)}{k^2} \right),
\label{Eq:DSEA}
\\
B(p^2)&=&Z_4 m + 4 \int_q \frac{\mathcal{G}(k^2)}{k^2} \frac{B(q^2)}{q^2A^2(q^2)+B^2(q^2)}.
\label{Eq:DSEB}
\end{eqnarray}
{where $k^\mu=p^\mu-q^\mu$.}The coupled Eqs.(\ref{Eq:DSEA}) and (\ref{Eq:DSEB}) can then be solved by direct iteration. The DSE solution for the dressed quark propagator can be well fitted with $3$ pairs of complex-conjugate poles with the representation~\cite{PhysRevD.67.054019,souchlas2010dressed,PhysRevD.70.014014,he2009model}
\begin{equation}
S(p)=\sum_{n=1}^3 \left( \frac{z_n}{i\slashed{p}+m_n} + \frac{z^\ast_n}{i\slashed{p}+m^\ast_n} \right).
\end{equation}
{Recently, this formula is generalized to study parton distribution amplitude of mesons~\cite{shi2014flavour,shi2015kaon}.}
We fit $u$, $d$ and $s$ quarks respectively, where the following requirement, that the quark propagator in the region of ultraviolet should tend to the free quark propagator, is employed
\begin{equation}
\sum_{k=1}^3 \left( z_k+z^\ast_k \right)=1.
\end{equation}
Corresponding parameters for $u$, $d$ quarks are
\begin{equation}
\begin{array}{ll}
m_1=349+111i\; \mathrm{MeV},	&z_1=0.3874+0.618i,\\
m_2=-1224+272i\; \mathrm{MeV},	&z_2=0.0912+0.176i,\\
m_3=1356+684i\; \mathrm{MeV},	&z_3=0.0214-0.097i.
\end{array}
\end{equation}
and for $s$ quark,
\begin{equation}
\begin{array}{ll}
m_1=714+215i\; \mathrm{MeV},	&z_1=0.4060+0.689i,\\
m_2=-1863-96i\; \mathrm{MeV},	&z_2=0.0987+0.0952i,\\
m_3=1429+893i\; \mathrm{MeV},	&z_3=-0.0047-0.093i.
\end{array}
\end{equation}

\section{Quark number density and EOS in $N_f=2$ and $N_f=2+1$ cases}\label{qndandpressure}
The solution of the quark propagator at $T=0$, $\mu=0$ case can be generalized to $T\neq 0$, $\mu\neq 0$ case by the following replacement~\cite{PhysRevC.71.015205}
\begin{equation}
p^4\rightarrow \tilde{\omega}_n=\omega_n+i\mu,
\end{equation}
which is widely used within rainbow truncation of DSEs in thermal and dense QCD, and ${\omega}_n=(2n+1)\pi T$ is the Matsubara frequencies. {It has been proved~\cite{PhysRevC.71.015205} that based on the rainbow truncation of the DSEs and two assumptions: (1) The full inverse quark propagator at finite chemical potential is analytic in the neighborhood of $\mu$= 0; (2) The effective gluon propagator is $\mu$ independent, the quark propagator at nonzero $T$ and $\mu$ can be obtained by replacement $p^4\rightarrow \tilde{\omega}_n=p^4+i\mu$.}
Hence, the quark propagator at nonzero $T$ and $\mu$ is
\begin{eqnarray}
S(\vec{p},\tilde{\omega}_n)&=&\sum_{k=1}^3 \bigg( \frac{z_k}{i\vec{\gamma}\cdot \vec{p}+i\gamma_4\tilde{\omega}_n+m_k}
\nonumber\\
&&\hspace*{5mm} + \frac{z^\ast_k}{i\vec{\gamma}\cdot \vec{p}+i\gamma_4\tilde{\omega}_n+m^\ast_k}  \bigg).
\label{quarkpg2}
\end{eqnarray}
The well-known formula for the quark number density is~\cite{PhysRevD.78.054001,he2007model}
\begin{equation}
\rho(\mu,T) = -N_c N_f\; T\!\!\sum_{n=-\infty}^{+\infty} \int \frac{d^3\vec{p}}{(2\pi)^3} \mathrm{tr}[S(\vec{p},\tilde{\omega}_n)\gamma_4].
\label{qnd1}
\end{equation}
Substituting Eq.(\ref{quarkpg2}) into Eq.(\ref{qnd1}), and then performing the trace, one obtains
\begin{eqnarray}
&&\rho(\mu,T)
\nonumber\\
&=& \sum_{k=1}^3 4i N_c N_f \int \frac{d^3\vec{p}}{(2\pi)^3} \;T\!\! \sum_{n=-\infty}^{+\infty} \bigg[ \frac{z_k(\omega_n+i\mu)}{\vec{p}^2+(\omega_n+i\mu)^2+m_k^2}
\nonumber\\
&&\hspace*{20mm}+ \frac{z^\ast_k(\omega_n+i\mu)}{\vec{p}^2+(\omega_n+i\mu)^2+{m^\ast_k}^2} \bigg]
\end{eqnarray}
The summation of infinite Matsubara frequencies can be evaluated using the method of contour integral~\cite{kapusta2006finite}
\begin{eqnarray}
&&T\sum_{n=-\infty}^{+\infty} f(p_0=i\omega_n)
\nonumber\\
&=& -\frac{1}{2} \sum_{\mathrm{Re}(p_0)>0} \mathrm{Res}\left[ f(p_0) \tanh\frac{p_0}{2T} + f(-p_0) \tanh\frac{p_0}{2T} \right]
\nonumber\\
\end{eqnarray}
where $p_0$ is regarded as a complex variable and $\mathrm{Res}[\cdots]$ means the residue of the function. Taking the limit $T\rightarrow 0$ of $\rho(\mu,T)$, one obtains
\begin{eqnarray}
&&\rho(\mu,T=0)
\nonumber\\
&=&\frac{N_c N_f}{3\pi^2} \sum_{k=1}^3 (z_k+z^\ast_k) \theta\left( \mu-\mu_k^0 \right) (\mu^2-\frac{d_k^2}{4\mu^2}-c_k)^{\frac{3}{2}}
\nonumber\\
\end{eqnarray}
where $\mu_k^0=|\mathrm{Re}(m_k)|$ and $c_k$, $d_k$ are defined by $m_k^2\equiv c_k+d_k i$.
\begin{figure}
\includegraphics[width=0.47\textwidth]{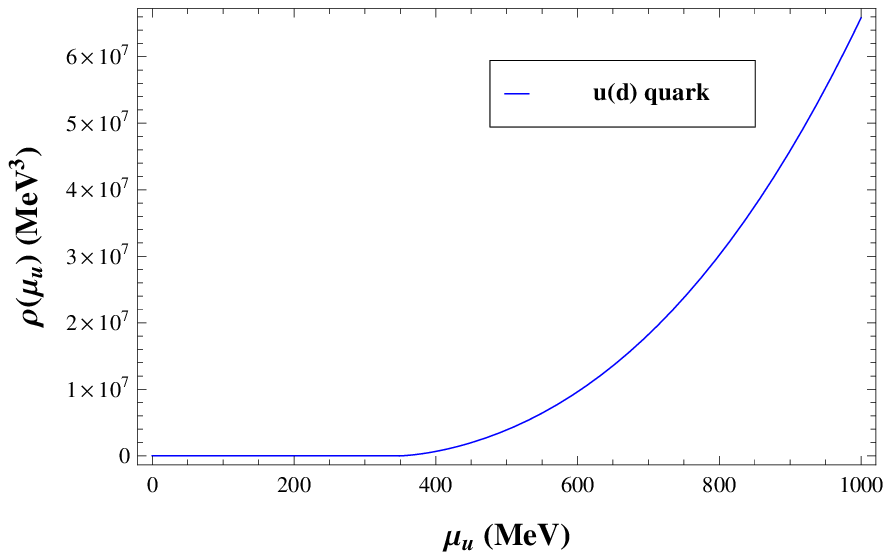}
\caption{Quark number density as a function of $\mu$ for $u$, $d$ quarks, where $T$ is taken to be 0.}
\label{Fig:qndu}
\end{figure}
\begin{figure}
\includegraphics[width=0.47\textwidth]{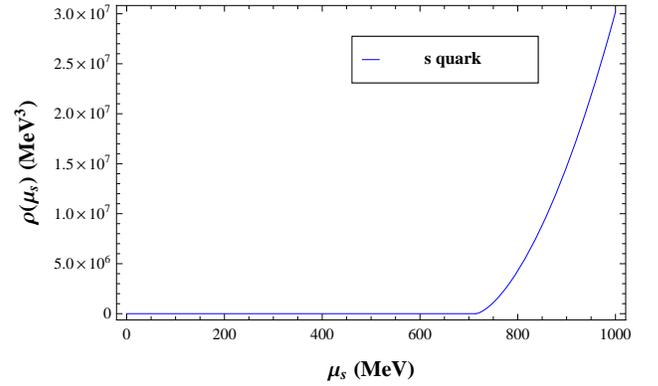}
\caption{Quark number density as a function of $\mu$ for $s$ quark, where $T$ is taken to be 0.}
\label{Fig:qnds}
\end{figure}

The quark number densities dependence on $\mu$ at $T=0$ for  $u$, $d$ and $s$ quarks are displayed in Fig.~\ref{Fig:qndu} and Fig.~\ref{Fig:qnds} respectively. $u$, $d$ quarks have the same quark number density dependence on their own chemical potential since the isospin symmetry is used and then their current quark mass are the same. We can see from Fig.~\ref{Fig:qndu} that there is a critical chemical potential $\mu_c=349$ MeV, at which the quark number density of $u$, $d$ quarks turn to nonzero from exact zero, while Fig.~\ref{Fig:qnds} shows that the quark number density of $s$ quark turns to nonzero at $\mu_c=714$ MeV.\footnote{In this work, the value of $\mu_c$ depends on the real part of the first mass pole.} This behavior agrees qualitatively with the general conclusion of~\cite{PhysRevD.58.096007}.

According to the electroweak theory of the Standard Model, 3 flavors of light quarks can transform with each other by the following reactions~\cite{PhysRevD.91.034018}.
\begin{eqnarray}
&d\rightarrow u+e^-+\bar{\nu}_e&,
\\
&u+e^-\rightarrow d+ \nu_e&,
\\
&s\rightarrow u+e^-+\bar{\nu}_e&,
\\
&u+e^-\rightarrow s+ \nu_e&,
\\
&u+d\leftrightarrow u+s&.
\end{eqnarray}
Since different quarks are coupled with each other, we can then employ the chemical equilibrium and electric charge neutrality conditions to constrain different chemical potentials, that is to say, $\mu_u$, $\mu_d$, $\mu_s$ and $\mu_e$. The conditions read
\begin{eqnarray}
&\mu_d=\mu_u+\mu_e,&\label{Eq:constrain1}
\\
&\mu_s=\mu_u+\mu_e,&\label{Eq:constrain2}
\\
&\frac{2}{3}\rho_u-\frac{1}{3}\rho_d-\frac{1}{3}\rho_s-\rho_e=0.&\label{Eq:constrain3}
\end{eqnarray}
There is only one independent chemical potential due to constraint equations Eqs.~(\ref{Eq:constrain1}), (\ref{Eq:constrain2}) and (\ref{Eq:constrain3}), we choose $\mu_u$ in this work. As discussed above, the quark number densities of $u$, $d$ and $s$ quarks at $T=0$ and finite $\mu$ are showed in Fig.~\ref{Fig:qndu} and Fig.~\ref{Fig:qnds}, and the particle number density of electron at $T=0$ is an explicit function of $\mu_e$, namely~\cite{kapusta2006finite}
\begin{equation}
\rho_e(\mu_e)=\frac{\mu_e^3}{3\pi^2}.
\end{equation}
For definite chemical potential of quark, the EOS of QCD at $T=0$ reads~\cite{he2007model,zong2008model}
\begin{equation}
P(\mu) = P(\mu=0) + \int_0^\mu d\mu^\prime \rho(\mu^\prime).
\end{equation}
In the case of 2 flavors, the EOS is just take $N_f=2$ for the two light quarks, $u$ and $d$, and the baryon number density is defined by
\begin{equation}
\rho^{2f}_B\equiv \frac{\rho_u(\mu_u) + \rho_d(\mu_d)}{3}.
\end{equation}
And in the case of 2+1 flavors, where $s$ quark is included, the pressure is
\begin{equation}
P(\mu_u)=P_u(\mu_u) + P_d(\mu_d) + P_s(\mu_s) + P_e(\mu_e),
\end{equation}
where $\mu_d,~\mu_s,~\mu_e$ are all related to $\mu_u$. The baryon number density in this case is then defined by
\begin{equation}
\rho^{3f}_B(\mu_u)\equiv\frac{\rho_u(\mu_u) + \rho_d(\mu_d) + \rho_s(\mu_s)}{3}.
\end{equation}
\begin{figure}
\includegraphics[width=0.47\textwidth]{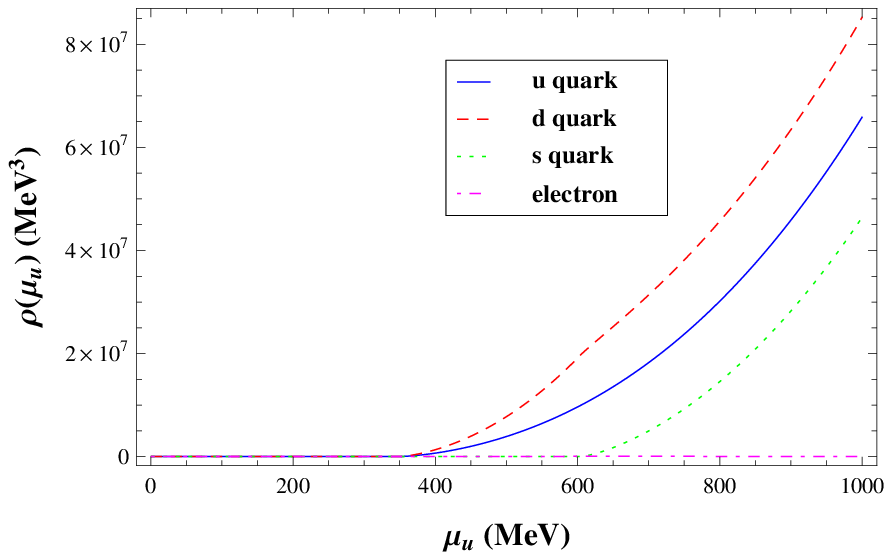}
\caption{The particle number density dependence of $u$, $d$, and $s$ quarks, as well as electron, on the chemical potential of $u$ quark, by the constrain of the chemical equilibrium and electric charge neutrality conditions.}
\label{Fig:qndwithmuu}
\end{figure}
\begin{figure}
\includegraphics[width=0.47\textwidth]{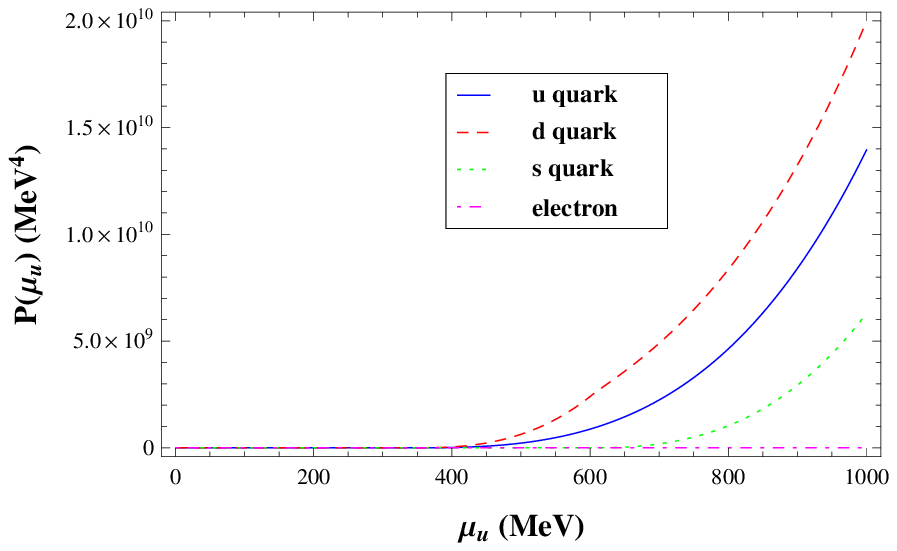}
\caption{The pressure dependence of $u$, $d$, and $s$ quarks, as well as electron, on the chemical potential of $u$ quark, by the constrain of the chemical equilibrium and electric charge neutrality conditions.}
\label{Fig:variouspressure}
\end{figure}
\begin{figure}
\includegraphics[width=0.47\textwidth]{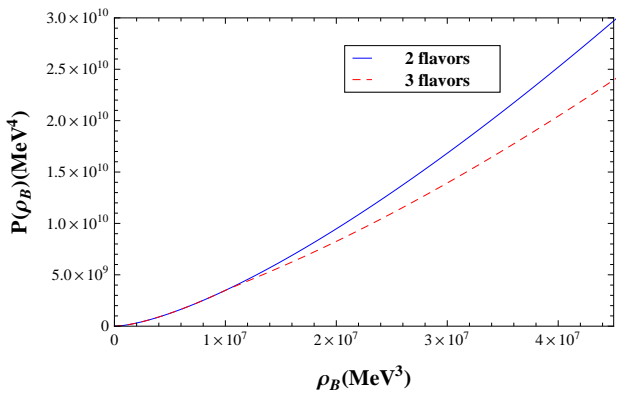}
\caption{The pressure dependence on baryon number density for $N_f=2$ and $N_f=3$ cases.}
\label{Fig:pressure}
\end{figure}
\begin{figure}
\includegraphics[width=0.47\textwidth]{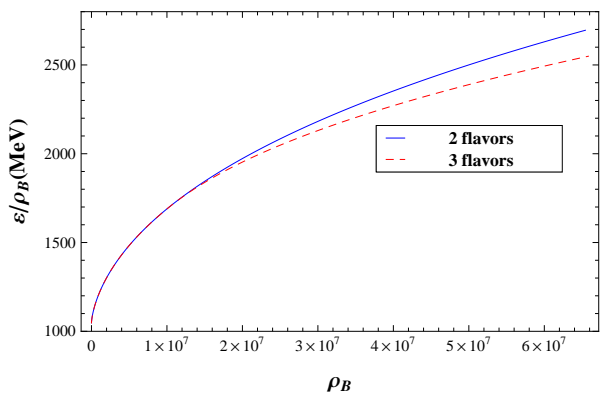}
\caption{The dependence of energy density per baryon on baryon number density for $N_f=2$ and $N_f=3$ cases.}
\label{Fig:energy}
\end{figure}
\begin{figure}
\includegraphics[width=0.47\textwidth]{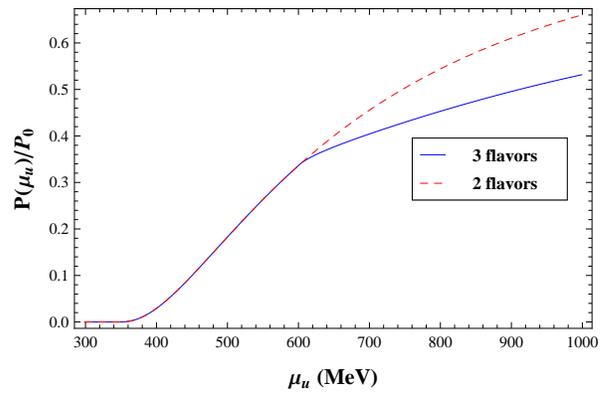}
\caption{The pressure compared with free quark gas for $N_f=2$ and $N_f=3$ cases.}
\label{Fig:EOSs}
\end{figure}
Now we plot the particle number density dependence and the pressure dependence of $u$, $d$, and $s$ quarks, as well as electron, on $u$ quark chemical potential $\mu_u$ in Fig. \ref{Fig:qndwithmuu} and Fig.~\ref{Fig:variouspressure} respectively. The pressure dependence of the system on baryon number density in cases of 2 flavors and 2+1 flavors are shown in Fig.~\ref{Fig:pressure}.
The quark number density of $u$ and $d$ quarks turn to nonzero at $\mu_u=349$ MeV with $\mu_u$ increasing. In the area of $\mu_u\lesssim 610$ MeV, the quark number density of $s$ quark is exact 0. It indicates that at this range the EOS and energy density of dense system has no difference between 2 flavors and 3 flavors cases. Combining with Fig. \ref{Fig:variouspressure}, we can see that the pressure of $u$, $d$, and $s$ quarks are $P_u=9.74\times 10^8~\mathrm{MeV}^4$, $P_d=2.65\times 10^9~\mathrm{MeV}^4$, and $P_s=0$ respectively.  We deduced from Fig. \ref{Fig:qndwithmuu} and Fig. \ref{Fig:variouspressure} that the $s$ quark has no effect on the pressure of the system when $\mu_u\lesssim 610$ MeV, and the critical chemical potential of $u$ quark, $\mu_u=610$ MeV, corresponds to baryon number density $\rho_B=1.027\times 10^7~\mathrm{MeV}^3$ and the pressure of system $P_T=3.62\times 10^9~\mathrm{MeV}^4$. Such conclusion is just depicted in Fig. \ref{Fig:pressure} that the two curves are indistinguishable when $\rho_B\lesssim 1.027\times 10^7~\mathrm{MeV}^3$. Two curves begin to split at $\rho_B= 1.027\times 10^7~\mathrm{MeV}^3$, namely $\mu_u= 610$ MeV, and the pressure in the case of 2+1 flavors is smaller than that in the case of 2 flavors at the area of $\rho_B\lesssim 1.027\times 10^7~\mathrm{MeV}^3$.

In the limit of large $\mu_u$, or in other words, large $\mu_u$, $\mu_d$, $\mu_s$ simultaneously, the behavior of pressure should tend to $N_c\mu_q^4/(12\pi^2)$ like the free quark gas, where $\mu_q$ means the chemical potential for $u$, $d$, or $s$ quarks. This can be easily understand since the quarks have the feature of asymptotic freedom. The pressure compared with free quark gas in 2+1 flavors and 2 flavors is depicted in Fig.~\ref{Fig:EOSs}. We can see from Fig.~\ref{Fig:EOSs} that the pressure of 2+1 flavors case tends to that of the free quark gas more slowly as the chemical potential increases. Compared with the work by Fraga $et~al$, our results imply that the interaction in the case of 2+1 flavors is stronger than 2 flavors case and that in the perturbative EOS~\cite{PhysRevD.63.121702,fraga2002new}. It is consistent with the view of strong coupled quark-gluon plasma, even in the case of large chemical potential.

The relation between the energy density and the pressure of the corresponding system is~\cite{PhysRevD.86.114028,PhysRevD.51.1989}
\begin{equation}
\varepsilon = -P + \sum_{i}\mu_i \rho_i,
\end{equation}
where $\mu_i$ and $\rho_i$ represents the chemical potential and the particle number density for each component in the system. We plot the dependence of energy density per baryon with the baryon number density in Fig. \ref{Fig:energy}. It is shown that the energy density per baryon is much smaller for the 2+1 flavors than that for the 2 flavors case when the $s$ quark number density is nonzero.
Our calculations imply that the system with 2+1 flavors of quarks is more stable than the case with 2 flavors of quarks.

\section{Discussion and Summary}\label{summary}
In this paper, the equation of state of 2+1 flavors of quarks, that is to say, $u$, $d$, and $s$, is studied within the framework of Dyson-Schwinger equations. We solve the DSE of 2+1 flavors quark propagator under the famous Rainbow-Ladder truncation, which preserves axial vector current conservation. An effective gluon propagator model, namely the Qin-Chang model~\cite{PhysRevC.84.042202}, is adopted. {It has been proved to describe hadron physics very well, especially for ground states. The gluon model generates DCSB, which is an important feature of low energy QCD. And it is infrared finite, which is consistent with lattice QCD's results~\cite{PhysRevD.80.085018,aguilar2010qcd,PhysRevC.84.042202}.} The parameters of this model, i.e., $m_u$, $m_d$, $D$, and $\omega$ are determined by fitting the masses and decay constants of pion and kaon. It is worth noting that these parameters are fixed during our study. The three complex-conjugate poles formula is used to fit the quark propagator at zero temperature and zero quark chemical potential. The quark propagator at finite temperature and finite chemical potential is generalized via the standard way in finite temperature field theory using the Matsubara frequencies. The chemical potential dependence of quark number densities are calculated at $T=0$. As Ref.~\cite{PhysRevD.58.096007} pointed out, the existence of some singularity at the point $\mu=\mu_c$, $T=0$ is a robust and model independent prediction, and the critical chemical potential for the lightest quarks ($u$ and $d$) is estimated to be $\mu=\mu_c\approx m_N-16$ MeV. In our study, the value of critical chemical potential is located at $\mu=349$ MeV, in which the quark number density turns to nonzero from 0 with the $u$ quark chemical potential increasing. It is qualitatively consistent with the general conclusion in~\cite{PhysRevD.58.096007}.
{As the quark chemical potential increases, the pressure of our result tends more slowly to the one of the free quark gas. This suggests that the interaction is stronger than the corresponding one in perturbation EOS, which is in consistent with the view of strongly coupled QGP, even in the case of large chemical potential.}

It is well known that different quarks can transform with each other by electroweak interactions, the chemical equilibrium and electric charge neutrality conditions are used to constrain the different chemical potentials, namely $\mu_u$, $\mu_d$, $\mu_s$ and $\mu_e$. We choose $\mu_u$ as the independent variable, in other words, $\mu_d$, $\mu_s$ and $\mu_e$ are all functions of $\mu_u$. In the cases of 2 flavors and 2+1 flavors, the quark number densities and pressure dependence on the chemical potential of $u$ quark are calculated, respectively. It is displayed in Fig.~\ref{Fig:qndwithmuu} that $\rho_s$ is exactly zero when $\mu_u<610$ MeV, so that the case with 2+1 flavors has no difference with the case of 2 flavors as showed in Fig. \ref{Fig:pressure}. While in the region of $\rho_B>1.027\times 10^7~\mathrm{MeV}^3$, the pressure in the case of 2+1 flavors is larger than the case with 2 flavors. Subsequently, the dependence of energy density per baryon on baryon number density for $N_f=2$ and $N_f=3$ cases is studied, Fig. \ref{Fig:energy} shows that the energy density per baryon with 2+1 flavors of quarks is smaller than that with 2 flavors. It implies that system with 2+1 flavors of quarks is more stable in the region where $\rho_B>1.027\times 10^7~\mathrm{MeV}^3$. Our results imply that the interaction in the case of three flavor is stronger than two flavors case and perturbation EOS.

Finally, we stress the two key points in this paper. Firstly, we generalize the EOSs in $SU_f(2)$ to that of $SU_f(3)$. Furthermore, We firstly combine DSEs with chemical equilibrium and  electric charge neutrality conditions to study EOS in dense QCD. These studies may provide some useful information to some research fields, such as the studies related to the QCD phase transitions or compact stars.

\acknowledgments
This work is supported in part by the National Natural Science Foundation of China (under Grant Nos. 11275097 and 11475085) and the Jiangsu Planned Projects for Postdoctoral Research Funds (under Grant No. 1402006C).
\bibliography{xuss}
\end{document}